**Курбанмурадов Давид Миколайович**
магістр
Державний університет телекомунікацій, Київ, Україна
ORCID ID: 0000-0003-4503-3773
*rockyou@protonmail.com*

**Соколов Володимир Юрійович**
старший викладач кафедри інформаційної та кібернетичної безпеки
Київський університет імені Бориса Грінченка, Київ, Україна
ORCID ID: 0000-0002-9349-7946
*v.sokolov@kubg.edu.ua*

**Астапеня Володимир Михайлович**
доцент кафедри інформаційної та кібернетичної безпеки
Київський університет імені Бориса Грінченка, Київ, Україна
ORCID ID: 0000-0003-0124-216X
*v.astapenia@kubg.edu.ua*

# РЕАЛІЗАЦІЯ ПРОТОКОЛУ ШИФРУВАННЯ XTEA НА БАЗІ БЕЗПРОВОДОВИХ СИСТЕМ СТАНДАРТУ IEEE 802.15.4

**Анотація.** У роботі вирішене завдання безпечності передачі даних в системах типу IEEE 802.15.4 на пристроях Pololu Wixel, наведені приклади апаратно-програмної реалізації шифрування і розшифрування різними пристроями на одній платформі. Запропоновані підходи можуть бути використані при розробці, впровадженні та експлуатації безпроводових корпоративних, промислових та персональних систем. Можливі напрямки розвитку цієї роботи пов'язані з виконанням досліджень щодо удосконалення алгоритмів шифрування (збільшення довжини ключа, використання асиметричних шифрів тощо), порівняння їх продуктивності та реалізація повноцінного протоколу обміну даними. В ході проведеної роботи виявилися проблеми в реалізації алгоритмів шифрування на малопотужних процесорах. Підчас роботи було вирішено ряд проблем, пов'язаних зі зведенням типів, адресацією, просторами пам'яті, переповненням буферу та іншими. Вирішені питання з узгодженням роботи приймача та передавача. Основне завдання роботи — побудова захищеного каналу зв'язку методом шифрування даних у каналі — було вирішено і отримана прошивка і прикладне програмне забезпечення для повноцінної перевірки роботи пристроїв. Також дана робота має великий прикладний потенціал, так як впровадження шифрування в існуючі системи незначним чином вплине на реалізацію і не вплине на кошторис проектів, але різко підвищить безпечність передавання даних в цих мережах. Запропоновані підходи можуть бути використані при розробці, впровадженні та експлуатації безпроводових корпоративних, промислових та персональних систем. Продовженням даної роботи може бути перевірка роботоспроможності інших протоколів на даному та аналогічному апаратному забезпеченні для систем, що можуть бути вбудовані в проекти обміну даними у безпроводових каналах зв'язку ближньої дії різних стандартів.

## 1. ВСТУП

Розвиток інтернету, корпоративних і приватних радіо мереж та Internet of Things (IoT) сильно вплинув на індустрію мереж та передачі даних. Одним із напрямків прогресу є радіомережі стандарту IEEE 802.15.4 Low-Rate Wireless Personal Area Networks (LR-WPANs) [1].

Стандарт IEEE 802.15.4 регламентує радіомережі повільної швидкості і малого радіусу дії (у порівнянні з IEEE 802.11 наприклад), які зазвичай використовуються в робототехніці, спеціалізованих корпоративних мережах та IoT [2, 3]. Головна мета цих пристроїв надати можливість підключати до них інші пристрої і обробляти і передавати їхні дані в тих чи інших цілях. Яскравим прикладом є так званий «розумний дім», який автоматизує та полегшує життя людини [4]. Оскільки в сферах де використовуються прилади даного типу циркулює конфіденційна інформація, то її треба належним чином захищати [5, 6]. В даній роботі було реалізовано блочний протокол шифрування даних, що надсилаються через радіоканал цих пристроїв.

## 2. ОСОБЛИВОСТІ ПРОТОКОЛУ ШИФРУВАННЯ XTEA

XTEA (eXtended TEA) — блоковий шифроалгоритм, створений усунути критичні помилки алгоритму TEA. Розробниками шифру є Девід Уілер і Роджер Нідхем з факультету комп'ютерних наук Кембриджського університету. Алгоритм був представлений в технічному звіті в 1997 році [7]. Шифр не запатентований, широко використовується в ряді криптографічних додатків і широкому спектрі апаратного забезпечення завдяки вкрай низьким вимогам до пам'яті і простоті реалізації [8].

Шифр заснований на операціях з 64-бітним блоком, має 32 повних цикли, в кожному повному циклі по два раунди мережі Фейстеля, що означає 64 раунди мережі Фейстеля з 128-бітним ключем. Однак, число раундів для досягнення кращої дифузії може бути збільшено на шкоду продуктивності. На рис. 1 показана спрощена схема шифрування ECB (Electronic Codebook), а на рис. 2 — CBC (Cipher Block Chaining) з вектором ініціалізації [9].

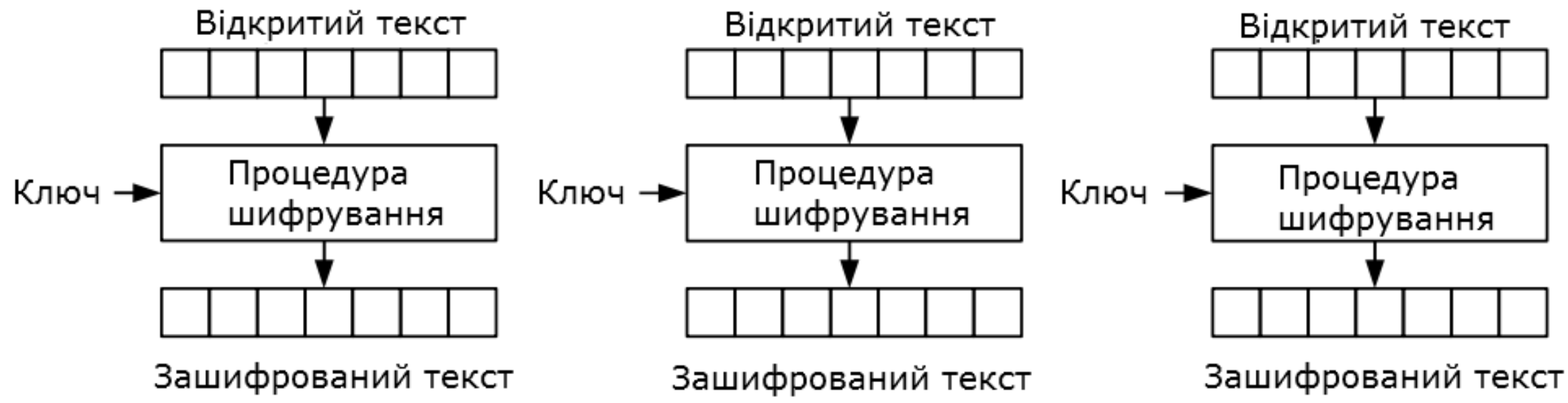

*Рис. 1. ECB режим шифрування*

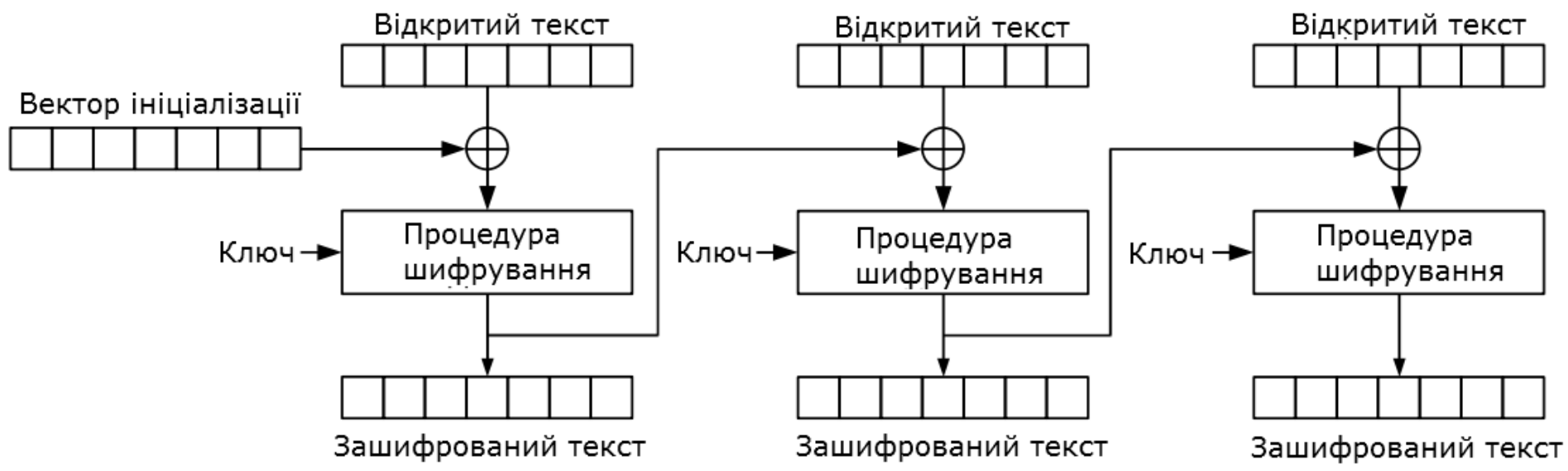

*Рис. 2. CBC режим шифрування*

## 3. ВИБІР АПАРАТНОГО ЗАБЕЗПЕЧЕННЯ ДЛЯ РЕАЛІЗАЦІЇ XTEA

Pololu Wixel — це програмований модуль загального призначення з радіо інтерфейсом, що працює на частоті 2,4 ГГц (табл. 1) [10, 11]. Використовує мікроконтролер TI CC2511F32. Призначений для використання у вбудованих системах та робототехніці. Розпіновка та пристрій зображені на зображенні на рис. 3. Мікроконтролер має наступні периферійні пристрої (табл. 2): 2 — USART, 3 — таймери сумісні з PWM, 1 —внутрішній таймер, 6 — аналогових GPIO, підключених до живлення у 7–12 В. Пристрій має 3 різнокольорових LED світлодіоди.

*Таблиця 1*

**Загальні характеристики Pololu Wixel**

| Характеристика | Значення |
|---|---|
| Процесор | CC2511-F32 @ 24 МГц |
| Розмір вбудованої пам’яті, кбайт | 4 |
| Розмір пам’яті програм, кбайт | 29 |
| Кількість користувацьких I/O ліній | 15 |
| Мінімальна напруга живлення, В | 2,7 |
| Максимальний напруга живлення, В | 6,5 |
| Запобігання переплюсовці | Так |
| Потреба в зовнішньому програматорі | Ні |

*Таблиця 2*

**Характеристики мікропроцесора CC2511-F32**

| Характеристика | Значення |
|---|---|
| Тип приладу | Wireless MCU |
| Розмір вбудованої пам’яті, кбайт | 24 |
| Мінімальна напруга живлення, В | 2 |
| Максимальний напруга живлення, В | 3,6 |
| Мінімальний струм живлення, мА | 14,7 |
| Максимальна швидкість передавання даних, кбіт/с | 500 |
| Види модуляцій | 2-FSK, GFSK, MSK |
| Максимальна чутливість, дБмВт | –103 |
| Максимальна потужність передавача, дБмВт | 1 |
| Наявність USB 2.0 | так |

В проекті використовується пристрій з мікроконтролером CC2511-F32, що відноситься до серії CC25XX [12]. Використовувана в них антенна типу Invented F Antenna (IFA) працює при 50 Ом і на частоті 2,45 ГГц. Пристрій має програмну можливість зміни радіоканалу для забезпечення стабільності, зменшення завад та розмежування пристроїв. Використовується протокол відповідно до IEEE 802.15.4.

Отже даних засобів та технологій достатньо, щоб створити наближену симуляцію інформаційної системи захищеного обміну повідомленнями всередині корпоративної безпроводової мережі.

## 4. ОСНОВНІ ЕЛЕМЕНТИ ЗАХИЩЕННОГО ОБМІНУ

Концепція захищеного обміну повідомленнями всередині корпоративних безпровідних мереж включає контроль доступу і цілісність повідомлень, конфіденційність і захист від повторів.

Контроль доступу означає, що протокол повинен запобігати участі неавторизованих сторін в мережі. Авторизовані вузли мають визначати повідомлення від неавторизованих сторін і відкидати їх. Крім того, захищена мережа повинна забезпечувати захист цілісності повідомлення якщо противник змінює повідомлення від авторизованого відправника поки повідомлення знаходиться в дорозі, приймач повинен бути в змозі виявити цю підробку. Включення Message Authentication Code (MAC) з кожним пакетом забезпечує автентифікацію і цілісність повідомлень. MAC може бути переглянутий як криптографічно захищена контрольна сума повідомлення. Вирахування її вимагає авторизованих відправників і одержувачів повідомляти секретний криптографічний ключ, і цей ключ є частиною вхідних даних для розрахунку. Відправник вираховує MAC по пакету за допомогою секретного ключа і включає MAC з пакетом. Отримувач з таким самим секретним ключем вираховує MAC пакету і порівнює його з тим, що в пакеті. Приймач приймає пакет, якщо вони рівні, і відхиляє його в іншому випадку. MAC дуже важко підробити без секретного ключа. Отже, якщо противник змінює дійсне повідомлення або робить ін'єкцію фальшивого повідомлення, він не зможе обчислити відповідний MAC, а також уповноважені приймачі будуть відкидати ці підроблені повідомлення, що потрапляють.

Конфіденційність означає зберігання інформації в таємниці від сторонніх осіб. Це, як правило, досягається за допомогою шифрування. Бажано, щоб шифрування не тільки захищало повідомлення від відновлення третіми сторонами, але також і попередити вивчення зловмисниками любої додаткової інформації про повідомлення, що були зашифровані. Такі дії відносяться до категорії семантичної стійкісті. Одним із аспектів семантичної стійкість є те, що шифрування одного й того ж повідомлення кілька раз в результаті має давати різний кінцевий результат. Якщо процес шифрування ідентичний для усіх звернень з однаковим повідомленням, тоді семантична стійкість явно порушується: результуючі шифротексти ідентичні. Загальним методом для досягнення семантичної безпеки є використання унікальної одноразової фрази (nonce) для кожного виклику алгоритму шифрування. Фразу можна розглядати як побічний вхід в алгоритм шифрування. Основною метою цієї фрази є варіювання процесу шифрування, коли є невелика зміна в наборі повідомлень. Так як приймач повинен використовувати цю фразу для розшифровки повідомлень (рис. 3), безпека більшості алгоритмів шифрування не вимагає щоб ці фрази були секретними, отже вони, як правило, посилаються у відкритому вигляді і включені в один пакет з зашифрованими даними.

Зловмисник, що підслуховує повідомлення що відправляються між авторизованими вузлами і повторює їх через якийсь час — виконує атаки повторного відтворення. Так як повідомлення походить від авторизованого відправника він буде мати дійсний MAC, так що приймач буде приймати його знову. Захист від повторів запобігає цим типам атак. Відправник зазвичай привласнює монотонно зростаючий порядковий номер до кожного пакету і приймач відхиляє пакети з меншими порядковими номерами, ніж він вже бачив.

| 1 байт | 2 байти | 1 байт | 0/2/4/10 байтів | 0/2/4/10 байтів | Змінна кількість | 2 байти |
|---|---|---|---|---|---|---|
| Довжина | Прапорці | Порядковий номер | Адреса отримувача | Адреса відправника | Корисне навантаження (дані) | Контрольна сума CRC |

*Рис. 3. Схема пакета*

Виконуючи всі ці умови можна досягти комплексної захищеності інформаційної системи захищеного обміну повідомленнями.

## 5. ПРОЕКТУВАННЯ КЛІЄНТСЬКОЇ ПРОГРАМИ

Програма-клієнт повинна:

– мати можливість відправки ключа для шифрування по USB на пристрої для заміни стандартного;

– приймати інформацію, що відправляють пристрої по USB;

– приймати інформацію від користувача, обробляти її та відправляти по USB на програму приймач.

Програма відправник повинна:

– мати можливість прийому ключа для шифрування по USB для заміни стандартного;

– приймати інформацію, що відправляє програма клієнт по USB, обробляти її, шифрувати, та відправляти по радіо в ефір;

– мати можливість відправляти дані по USB.

Програма-приймач повинна:

– мати можливість прийому ключа для шифрування по USB для заміни стандартного;

– приймати інформацію, що відправляє програма відправник в радіо ефір, розшифрувати її, обробляти, та відправляти по USB;

– мати можливість відправляти дані по радіо.

## 6. РЕАЛІЗАЦІЯ ПРОТОКОЛУ ШИФРУВАННЯ

В рамках даного проекту використовується алгоритм блочного шифрування XTEA на мові програмування C з компілятором SDCC (див. лістинги 1 і 2 в двох режимах роботи) [13].

*Лістинг 1. Функція шифрування/розшифрування ECB режим*

```
int mbedtls_xtea_crypt_ecb( mbedtls_xtea_context *ctx, int mode,
const unsigned char input[8], unsigned char output[8]) {
  uint32_t *k, v0, v1, i;
  k = ctx->k;
  GET_UINT32_BE( v0, input, 0 );
  GET_UINT32_BE( v1, input, 4 );
  if( mode == MBEDTLS_XTEA_ENCRYPT ) {
    uint32_t sum = 0, delta = 0x9E3779B9;
    for( i = 0; i < 32; i++ ) {
      v0 += (((v1 << 4) ^ (v1 >> 5)) + v1) ^ (sum + k[sum & 3]);
      sum += delta;
      v1 += (((v0 << 4) ^ (v0 >> 5)) + v0) ^ (sum + k[(sum>>11) & 3]);
    }
  } else {
    uint32_t delta = 0x9E3779B9, sum = delta * 32;
    for( i = 0; i < 32; i++ ) {
      v1 -= (((v0 << 4) ^ (v0 >> 5)) + v0) ^ (sum + k[(sum>>11) & 3]);
      sum -= delta;
      v0 -= (((v1 << 4) ^ (v1 >> 5)) + v1) ^ (sum + k[sum & 3]);
    }
  }
  PUT_UINT32_BE( v0, output, 0 );
  PUT_UINT32_BE( v1, output, 4 );
  return( 0 );
}
```

*Лістинг 2. Функція шифрування/розшифрування CBC режим*

```
     int mbedtls_xtea_crypt_cbc( mbedtls_xtea_context *ctx, int mode, size_t length,
unsigned char iv[8], const unsigned char *input, unsigned char *output) {
  int i;
  unsigned char temp[8];
  if( length % 8 )
    return( MBEDTLS_ERR_XTEA_INVALID_INPUT_LENGTH );
  if( mode == MBEDTLS_XTEA_DECRYPT ) {
    while( length > 0 ) {
      memcpy( temp, input, 8 );
      mbedtls_xtea_crypt_ecb( ctx, mode, input, output );
      for( i = 0; i < 8; i++ )
        output[i] = (unsigned char)( output[i] ^ iv[i] );
      memcpy( iv, temp, 8 );
      input  += 8;
      output += 8;
      length -= 8;
    }
  } else {
    while( length > 0 ) {
      for( i = 0; i < 8; i++ )
        output[i] = (unsigned char)( input[i] ^ iv[i] );
        mbedtls_xtea_crypt_ecb( ctx, mode, output, output );
        memcpy( iv, output, 8 );
        input  += 8;
        output += 8;
        length -= 8;
      }
  }
  return( 0 );
}
```

В даному проекті розглядається симуляція спрощеної інформаційної системи обміну захищеними повідомленнями для визначення можливостей цієї системи з даними типами пристроїв та перспективами їх розвитку.

## 7. ІНТЕГРАЦІЯ З ПРИСТРОЄМ І ВПРОВАДЖЕННЯ

Виходячи з апаратних та програмних особливостей пристроїв Pololu Wixel при виконанні даного проекту було виявлено наступні проблеми:

– зведення типів;
– переповнення буферу;
– робота з вказівниками;
– змінні в пам’яті XDATA;
– зациклювання з мінімальною затримкою;
– помилки компілятора.

Зведення типів. Через те, що використовувана в SDCC мова програмування C є мовою програмування з жорсткою типізацією, виникла проблема зведення типів. Наприклад компілятор забороняє прирівнювати *uint8* та *char*, хоча *char* потребує для себе також 1 байт. Також не можна прирівнювати між собою визначені масиви. Звідси виникає необхідність зведення типів даних між змінними та навмисного вказування зведення при передачі параметра функції. Зведення типів, в більшості випадків, не дає бажаного результату, бо мова програмування не передбачає конвертування тих чи інших типів даних.

Рішенням може бути один з варіантів:

1. Використовувати однакові типи даних.
2. Використовувати функцію *sprintf()* з бібліотеки *stdio* (при цьому треба перевизначити функцію *putchar()* оскільки у приладу немає стандартного методу виводу і необхідно його визначити).

```
void putchar(char c) { usbComTxSendByte(c); }}
```

**Переповнення буферу.** Частою проблемою стало переповнення буферу. Оскільки прилад має дуже обмежену пам’ять і призначений для вбудовуваних систем то треба слідкувати за всіма створеними змінними і робити код таким, щоб одночасно не використовувалось дуже багато змінних. Також не бажано використовувати масиви з невизначеною довжиною(наприклад *uint16 XDATA **) для зберігання в них динамічних елементів(довжина яких може коливатися), бо це може призвести втрату даних чи переповнення буферу. При переповненні буферу прилад перестає приймати і відправляти та стає некерований, в такому разі вимкнути його можна лише зробивши «гаряче» від’єднання. Створювати змінні в пам’яті XDATA та CODE бажано лише при необхідності через те, що для кожної такої змінної виділяється 64 кілобайт простору пам’яті. Рішення: слідкувати за змінними, їх кількістю, об’ємами та просторах пам’яті.

**Робота з вказівниками.** Через те, що багато функцій в API Wixel SDK приймають в якості параметрів змінні-масиви з невизначеною довжиною доводиться їх використовувати та/або зводити до них інші типи, що спричиняє вище описані проблеми а також втрату даних. Наприклад, вбудована в API функція usbComTxSend об’явлена наступним чином:

```
void usbComTxSend(const uint8 XDATA * buffer, uint8 size);
```

Використовувати такі масиви тільки в функціях з API та перетворювати їх до інших типів за допомогою функції *sprintf()*.

**Змінні в пам’яті XDATA.** Як вище вже описувалося, розмір кожної змінної в просторі XDATA може досягати 64 кілобайт. В умовах того що таких змінних створюється багато прилад перестає виконувати програму та переходить в bootloader-режим (стан 0×0100). Рішення: не використовувати багато змінних в просторі XDATA.

**Зациклювання з мінімальною затримкою.** Для того щоб програма працювала весь час доки прилад увімкнено використовується безкінечний цикл *while(1)*. Ініціалізація відбувається до нього, а основні функції визиваються всередині нього. Виходячи з цього, якщо функції всередині виконуються швидко, то забивається буфер COM-порту або ж радіоінтерфейсу прийомної сторони і інша програма (девайс) не встигає оброблювати всі дані. Рішення: використовувати затримку програми за допомогою вбудованої функції delayMs().

**Помилки компілятора.** Коли в рамках проекту використовувалася IDE Eclipse for C/C++ компілятор записував програму не за тією адресою пам’яті і перезаписував системні зони, таким чином прилади ставали непрацездатними. Скоріш за все це пов’язано з тим, що IDE може підключати до проекту свої бібліотеки примусово замінивши бібліотеки SDCC. Рішення: відмовитись від використання IDE та використовувати програмне забезпечення надане Pololu Wixel для розробки. Порядок відновлення приладів наступним чином:

– замикається P2_2 та 3V3;
– підключається до комп’ютера;
– прошивається за допомогою Pololu Wixel Configuration Utility;
– відключається;
– знімається замикаючий інструмент;
– підключається.

Користувач вводить дані в клієнтську програму, програма обробляє та відправляє ці дані відправникові через USB (рис. 4). За необхідності програма відправляє ключі відправникові та приймачу перед тим як відправити оброблені дані. Відправник шифрує отримані дані, додає свою унікальну MAC-адресу та відправляє все це по радіоканалу в ефір. Приймач отримує з ефіру надіслані дані розшифровує їх та передає в клієнтську програму по USB. Відправник і приймач мають можливість передачі відкритих даних по USB.

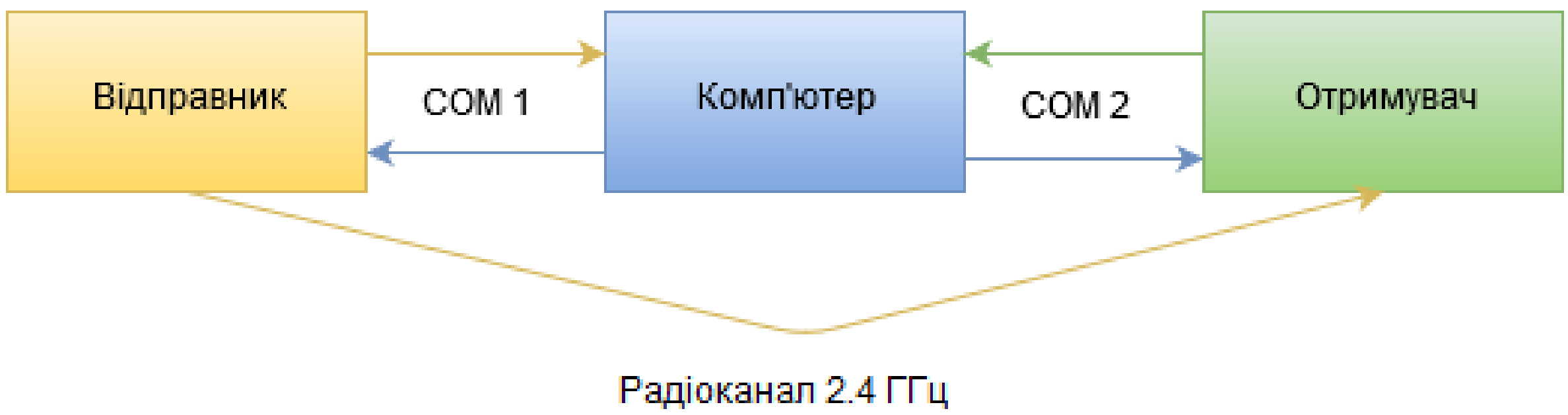

*Рис. 4. Спрощена схема даних проекту*

Для ініціалізації з’єднання між комп’ютером та приладами достатньо знати, який COM-порт прилад використовує та швидкість з’єднання, яку він підтримує (лістинг 3).

*Лістинг 3. Ініціалізація з'єднання з COM*

```
System.IO.Ports.SerialPort tp = new System.IO.Ports.SerialPort("COM3", 115200);
tp.Open();
```

Для того щоб прилад зрозумів що зараз надсилатиметься ключ (лістинг 4), який треба буде використати замість стандартного, надсилаємо йому групу з восьми однакових байт що містять символ управління ENQUIRY (номер 05) з таблиці ASCII (лістинг 5), який призначений для використання в телетайпному зв'язку і в даній ситуації може використовуватись у своїх цілях (рис. 5).

*Лістинг 4. Передавання ключа на пристрій*

```
string key1 = "a1b2c3d4", key2 = "fffyfff0";
byte[] b0 = new byte[] {0x05, 0x05, 0x05, 0x05, 0x05, 0x05, 0x05, 0x05};
byte[] b1 = Encoding.Default.GetBytes(key1);
byte[] b2 = Encoding.Default.GetBytes(key2);
tp.Write(b0, 0, 8);
tp.Write(b1, 0, 8);
tp.Write(b2, 0, 8);
```

*Лістинг 5. Перевірка даних на наявність символу управління 05 на боці приладу*

```
if(buffer[0]==05 && buffer[7]==05) {
  printf("#ok\r\n");
  isakey=1;
}
```

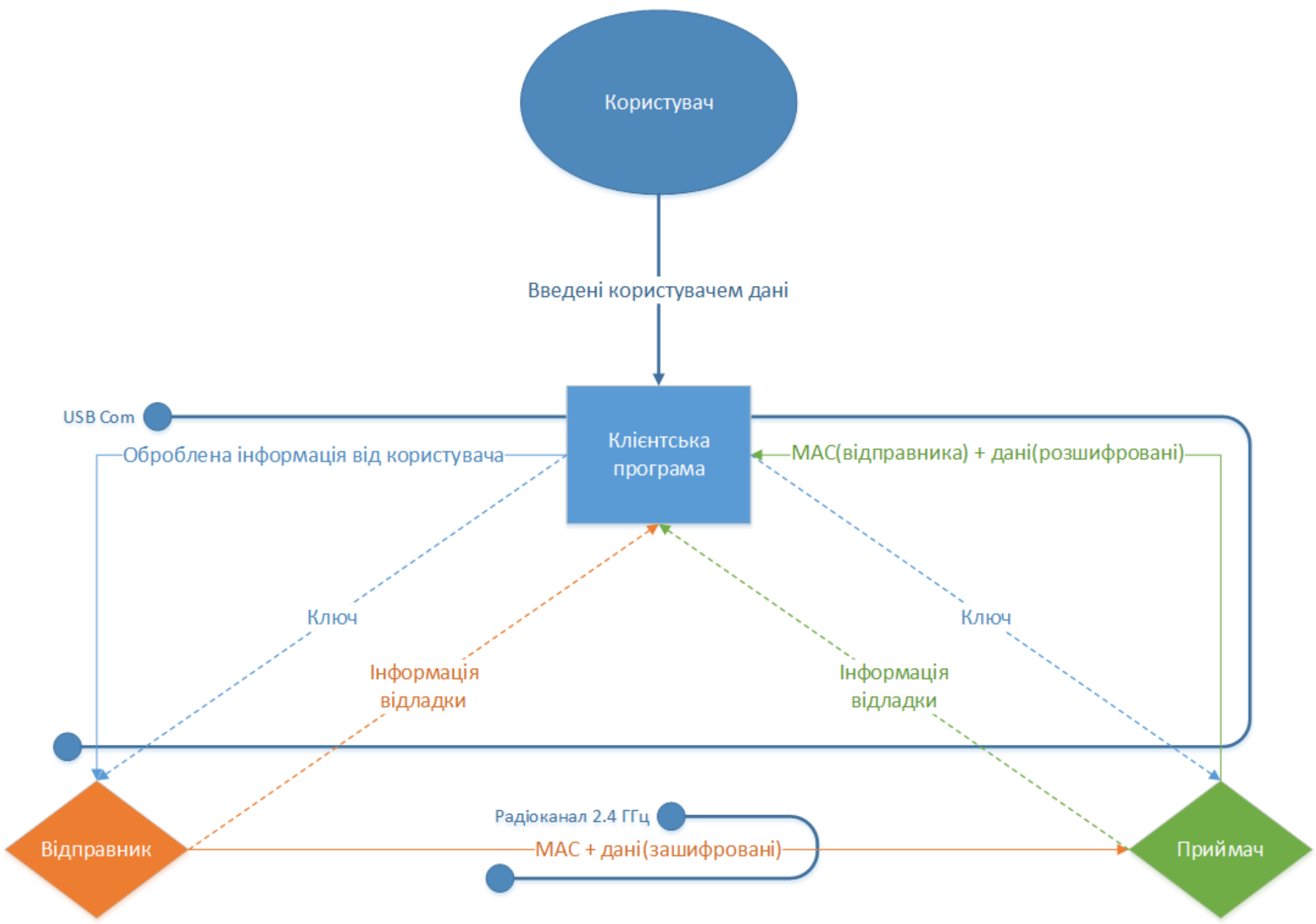

*Рис. 5. Схема даних проекту*

При розділення даних на стороні комп'ютера використовуються: кількість ітерацій (*it*), залишок на останній ітерації (*lst*) і кількість елементів яких не вистачає до заповнення останньої ітерації (*nul*), які приведені на лістингу 6.

*Лістинг 6. Функція на стороні комп'ютера, що розподіляє введені дані*

```
string b = Console.ReadLine();
int it, lst = b.Length, nul = 0;
if (b.Length % 8 != 0) {
  it = (b.Length / 8) + 1;
} else {
  it = (b.Length / 8);
}
while (lst > 8) { lst = lst - 8; }
for (int i = 0; i < it; i++) {
  int len = 8;
  if (i == it - 1) {
    len = lst;
    nul = 8 - lst;
  }
  byte[] bytes = Encoding.Default.GetBytes(b.Substring(i * 8, len));
  if (nul > 0 ) {
    Array.Resize(ref bytes, bytes.Length + nul);
  }
  tp.Write(bytes,0,bytes.Length);
  Thread.Sleep(20);
}
```

Спочатку перевіряємо чи кратна довжина введених даних восьми, звідси визначаємо скільки ітерацій нам потрібно провести (лістинг 7). Потім отримуємо скільки елементів залишається на останній (можливо неповній) ітерації. Після чого вираховуємо кількість елементів яких недостає і збільшуємо масив на цю кількість. При зміні розміру масиву .NET автоматично ініціює всі невизначені елементи нулями.

*Лістинг 7. Функція шифрування/розшифрування на приладах*

```
uint8 * encrypt(uint8 * input, int type) {
  uint8 tmp[8];
  uint8 output[8];
  mbedtls_xtea_context * enc = NULL;
  sprintf(tmp, input);
  mbedtls_xtea_init(enc);
  mbedtls_xtea_setup(enc, key);
  mbedtls_xtea_crypt_ecb(enc, type, tmp, output);
  mbedtls_xtea_free(enc);
  return output;
}
```

На вході приймає вказівник на масив з даними, тобто достатньо передати назву масиву, оскільки вона є вказівником на перший (0) елемент масиву, та режим обробки (кодування — 0 чи декодування — 1). На виході масив з результатом обробки показано на лістингу 8.

*Лістинг 8. Формування пакета та відправка його в ефір*

```
uint8 XDATA * txPacket;
txPacket = radioQueueTxCurrentPacket();
uint8 XDATA * ptr = (uint8 XDATA *)&txPacket[5];
```

```
txPacket[0] = 16;
txPacket[1] = serialNumber[0];
txPacket[2] = serialNumber[1];
txPacket[3] = serialNumber[2];
txPacket[4] = serialNumber[3];
sprintf(rinput, buffer);
sprintf(ptr, encrypt(rinput, 0));
radioQueueTxSendPacket();
```

Розшифрування та надсилання на комп'ютер показано на лістингу 9.

*Лістинг 9. Розшифрування та відправлення пакету*

```
sprintf(routput, encrypt(rinput, 1));
for(i = 0; i < 8; i++) {
      printf("%c", routput[i]);
}
```

Ми використовуємо функції *sprint* і *printf* для конвертування між різними типами і надсилання даних по USB відповідно (лістинг 10).

*Лістинг 10. Вивід прийнятих даних на комп'ютері*

```
string x = tp.ReadExisting();
Console.WriteLine(x);
```

На комп'ютері отримувати дані можна нерозподілено, оскільки його пам'яті набагато більше ніж пам'яті приладів [14]. Загалом інтеграція пройшла успішно з деякими побічними проблемами, що виникали на протязі реалізації проекту на використовувану платформу, які були ідентифіковані, досліджені та вирішені. Пристрої показали себе як дуже перспективні і витривалі, отже даний проект буде продовжений з новими цілями і завданнями.

## 8. ВИСНОВКИ ТА ПЕРСПЕКТИВИ ПОДАЛЬШИХ ДОСЛІДЖЕНЬ

В ході проведеної роботи виявилися проблеми в реалізації алгоритмів шифрування на малопотужних процесорах. Через брак ресурсів пристрою був вибраний блочний протокол шифрування XTEA, який хоча ще не має широкого визнання, але показав гарні результати у нашій конкретній задачі. Підчас роботи було вирішено ряд проблем, пов'язаних зі зведенням типів, адресацією, просторами пам'яті, переповненням буферу та іншими. Вирішені питання з узгодженням роботи приймача та передавача. Основне завдання роботи — побудова захищеного каналу зв'язку методом шифрування даних у каналі — було вирішено і отримана прошивка і прикладне програмне забезпечення для повноцінної перевірки роботи пристроїв. Також дана робота має великий прикладний потенціал, так як впровадження шифрування в існуючі системи незначним чином вплине на реалізацію і не вплине на кошторис проектів, але різко підвищить безпечність передавання даних в цих мережах.

Продовженням даної роботи може бути перевірка роботоспроможності інших протоколів на даному та аналогічному апаратному забезпеченні для систем, що можуть бути вбудовані в проекти обміну даними у безпроводових каналах зв'язку ближньої дії. А також окремий інтерес має визначення залежності інформаційної швидкості передавання даних від довжини ключа. В майбутньому планується зробити перевірку пакетів та впровадити алгоритм отримання втрачених пакетів; перевести клієнтську частину на Python для можливості кросплатформенного використання; зробити одну універсальну прошивку для всіх приладів (TX і RX); передбачити можливість використання багатьох (більше 2-х) приладів на одному каналі; впровадити, протестувати і дослідити наступні алгоритми шифрування: AES, ARCFOUR, Blowfish / BF, Camellia, DES/3DES, GCM, Diffie-Hellman-Merkle, RSA, Elliptic Curves over GF (p), Elliptic Curve Digital Signature Algorithm (ECDSA), Elliptic Curve Diffie Hellman (ECDH).

## СПИСОК ВИКОРИСТАНИХ ДЖЕРЕЛ

**Davyd M. Kurbanmuradov**
MSc
State University of Telecommunications, Kyiv, Ukraine
ORCID ID: 0000-0003-4503-3773
*rockyou@protonmail.com*

**Volodymyr Yu. Sokolov**
MSc, senior lecturer
Borys Grinchenko Kyiv University, Kyiv, Ukraine
ORCID ID: 0000-0002-9349-7946
*v.sokolov@kubg.edu.ua*

**Volodymyr M. Astapenya**
PhD, associate professor
Borys Grinchenko Kyiv University, Kyiv, Ukraine
ORCID ID: 0000-0003-0124-216X
*v.astapenia@kubg.edu.ua*

# IMPLEMENTATION OF XTEA ENCRYPTION PROTOCOL BASED ON IEEE 802.15.4 WIRELESS SYSTEMS

**Abstract.** The problem of data security in IEEE 802.15.4 systems on Pololu Wixel devices is solved, examples of hardware and software implementation of encryption and decryption of different devices on the same platform are given. The proposed approaches can be used in the development, implementation and operation of wireless enterprise, industrial, and personal systems. Possible areas of development for this work are related to research on improving encryption algorithms (increasing key length, using asymmetric ciphers, etc.), comparing their performance, and implementing a complete data exchange protocol. During the work there were problems in the implementation of encryption algorithms on low-power processors. During the work, a number of issues were resolved regarding type reduction, addressing, memory space, buffer overflow, and more. Issues resolved with reconciliation of receiver and transmitter operation. Examples of hardware and software implementation of encryption and decryption of different devices based on Pololu Wixel are given in the paper. The basic task of building a secure communication channel by encrypting data in the channel was solved and firmware and application software were obtained to fully validate the devices. In addition, this work has great application potential, since the implementation of encryption in existing systems will have a small impact on implementation and will not affect the project budget, but will dramatically improve the security of data transmission in these networks. The proposed approaches can be used in the development, implementation and operation of wireless enterprise, industrial, and personal systems. Continuing this work may be to test the performance of other protocols on this and similar hardware for systems that may be embedded in short-range wireless communication projects of short-range standards.